\documentclass[10pt,reqno]{amsart}

\usepackage{epsfig}
\usepackage{amsmath,amssymb}

\numberwithin{equation}{section}

\newcommand{\wt}{\widetilde}

\newcommand{\R}{{\mathbb R}}

\renewcommand{\phi}{\varphi}

\newcommand{\hcal}{\mathcal{H}}

\newcommand{\be}{\beta}
\newcommand{\ga}{\gamma}

\newcommand{\La}{\Lambda}
\newcommand{\la}{\lambda}
\newcommand{\ep}{\varepsilon}
\newcommand{\eps}{\epsilon}
\newcommand{\de}{\delta}
\newcommand{\De}{\Delta}
\newcommand{\f}{\varphi}

\newcommand{\DT}{\Delta T}
\newtheorem{theo}{{\sc Theorem}}[section]

\newenvironment{rem}{\medskip\noindent{\it Remark:\/} }{\medskip}

\renewcommand{\appendix}[1]{
    \addtocounter{section}{1}
    \setcounter{equation}{0}
    \renewcommand{\thesection}{\Alph{section}}
    \section*{Appendix \thesection\protect\indent #1}
    \addcontentsline{toc}{section}{Appendix \thesection\ \ \ #1}
}
\newcommand\encadremath[1]{\vbox{\hrule\hbox{\vrule\kern8pt 
\vbox{\kern8pt \hbox{$\displaystyle #1$}\kern8pt} 
\kern8pt\vrule}\hrule}}
\def\enca#1{\vbox{\hrule\hbox{
\vrule\kern8pt\vbox{\kern8pt \hbox{$\displaystyle #1$}
\kern8pt} \kern8pt\vrule}\hrule}}

\newcommand\figureframex[3]{
\begin{figure}[bth]
\hrule\hbox{\vrule\kern8pt 
\vbox{\kern8pt \vbox{
\begin{center}
{\mbox{\epsfxsize=#1.truecm\epsfbox{#2}}}
\end{center}
\caption{#3}
}\kern8pt} 
\kern8pt\vrule}\hrule
\end{figure}
}
\newcommand\figureframey[3]{
\begin{figure}[bth]
\hrule\hbox{\vrule\kern8pt 
\vbox{\kern8pt \vbox{
\begin{center}
{\mbox{\epsfysize=#1.truecm\epsfbox{#2}}}
\end{center}
\caption{#3}
}\kern8pt} 
\kern8pt\vrule}\hrule
\end{figure}
}

\newcommand{\beq}{\begin{equation}}
\newcommand{\eeq}{\end{equation}}
\newcommand{\bea}{\begin{eqnarray}}
\newcommand{\eea}{\end{eqnarray}}

\renewcommand{\and}{{\qquad {\rm and} \qquad}}

\newcommand{\virg}{{\qquad , \qquad}}

 \newcommand{\Tr}{{\,\rm Tr}\:}
\newcommand{\om}{\omega}

\newcommand{\Pint}{{\int\kern -1.em -\kern-.25em}}

\title[Double Scaling Limit in Random Matrix Models]
{Double Scaling Limit in Random Matrix Models  
and a Nonlinear Hierarchy of Differential Equations}

\author{Pavel Bleher}

\thanks{Research of the first author (P.B.)
was partially supported by NSF grant
\#DMS-9970625. The second author (B.E.)
 would like to thank the CRM, Prof. J. Harnad and Prof. 
J. Hurtubise for their support when this work was completed.}

\address{Department of Mathematical Sciences\\
Indiana University-Purdue University Indianapolis \\ 
402 N. Blackford Street, Indianapolis, IN
46202, USA}

\email {bleher@math.iupui.edu}

\author{Bertrand Eynard}

\address{Service de Physique Th\'{e}orique, Saclay\\
F-91191 Gif-sur-Yvette Cedex, France\\
and  Centre de recherches math\'ematiques\\
Universit\'e de Montr\'eal\\ C.~P.~6128\\ succ. centre ville, Montr\'eal\\
Qu\'ebec, Canada H3C 3J7}

\email {eynard@spht.saclay.cea.fr}

\date{\today}

\begin{document}

\begin{abstract} We derive the double scaling limit of eigenvalue correlations in 
  the random matrix model at critical points
and  we relate it to a nonlinear
hierarchy of ordinary differential equations.  
\end{abstract}

\maketitle

\section {Introduction}

We consider the unitary ensemble of random matrices,
\begin{equation}\label{qi1}
d\mu_N(M)=Z_N^{-1}\exp\left(-N\Tr V(M)\right) dM,\quad
Z_N=\int_{{\hcal}_N} 
\exp\left(-N\Tr V(M)\right) dM,  
\end{equation}
on the space ${\hcal}_N$
of Hermitian $N\times N$ matrices $M=\left(M_{ij}\right)_{1\le
i,j\le N}$, where $V(x)$ is a polynomial, $V(x)=v_px^p+v_{p-1}x^{p-1}+\ldots$,
of an even degree $p$ with $v_p>0$.
The ensemble of eigenvalues $\la
=\{\la_j,\;j=1,\ldots,N\}$ of $M$ is given then by the formula
(see e.g. \cite{Meh}, \cite{TW}),
\begin{equation}\label{qi1r}
d\mu_N(\la)=\wt Z_N^{-1}\exp\left(-NH_N(\la)\right) d\la,\quad
\wt Z_N=\int_{\La_N} 
\exp\left(-NH_N(\la)\right) d\la,  
\end{equation}
where $\La_N$ is the symmetrized $\R^N$, $\La_N=\R^N/S(N)$,
 and
\begin{equation}\label{qi1b}
H_N(\la)=-\frac{2}{N}\,\sum_{1\le j<k\le N}\log|\la_j-\la_k|+
\sum_{j=1}^N V(\la_j)\,.  
\end{equation}
Let $d\nu_N(x)=\rho_N(x)dx$
be the distribution of the eigenvalues
on the line,
so that for any test function $\phi(x)\in C^{\infty}_0$,
\begin{equation}\label{qi1c}
\int_{\La_N}\left[ \frac{1}{N}\sum_{j=1}^N \phi(\la_j)\right]\,
d\mu_N(\la)
=\int_{-\infty}^\infty \phi(x)\,d\nu_N(x)\,.
\end{equation}
  As $N\to\infty$, there exists a weak limit of $d\nu_N(x)$,
\begin{equation}\label{qi1d}
d\nu_{\infty}(x)=\lim_{N\to\infty} d\nu_N(x)\,.
\end{equation}
To determine the limit (cf. \cite{BIPZ}, \cite{DGZ},
and others),
consider the energy functional on the space of probability measures
on the line,
\begin{equation}\label{qi1e}
I(d\nu(x))=-\iint_{\R^2}\log|x-y|d\nu(x)\,d\nu(y)+\int_{\R}V(x)d\nu(y)\,.
\end{equation}
Then $H_N(\la)$ in (\ref{qi1b}) can be written as
\begin{equation}\label{qi1f}
H_N(\la)=NI(d\nu(x;\la))\,,
\end{equation}
where $d\nu(x;\la)$ is a discrete probability measure with atoms at $\la_j$,
\begin{equation}\label{qi1g}
d\nu(x;\la)=\frac{1}{N}\sum_{j=1}^N\de(x-\la_j)\,dx\,.
\end{equation}
Hence,
\begin{equation}\label{qi1h}
d\mu_N(\la)=\wt Z_N^{-1}\exp\left(-N^2 I(d\nu(x;\la))\right) d\la\,.
\end{equation}
Because of the factor $N^2$ in the exponent, one can expect that 
as $N\to\infty$, the measures $d\mu_N(\la)$ are localized in a shrinking
vicinity of an {\it equilibrium measure} $d\nu_{\rm eq}(x)$, which
minimizes the functional $I(d\nu(x))$, and therefore, one expects the limit
(\ref{qi1d}) to exist with $d\nu_{\infty}(x)=d\nu_{\rm eq}(x)$.
A rigorous
proof of the existence and uniqueness of the equilibrium measure,
its properties,
and the existence of limit (\ref{qi1d}) 
with $d\nu_{\infty}(x)=d\nu_{\rm eq}(x)$,
was given in \cite{BPS} and \cite{Joh}.

The equilibrium measure $d\nu_{\rm eq}(x)$ 
is supported by a finite number of 
segments $[a_j,b_j]$, $j=1,\ldots,q$, and it is absolutely continuous
with respect to the Lebesgue measure, $d\nu_{\rm eq}(x)=\rho(x)dx$,
with a density function $\rho(x)$ of the form
\begin{equation}\label{qi2}
\rho(x)=\frac{1}{2\pi i} h(x)R_+^{1/2}(x),\quad
R(x)=\prod_{j=1}^q(x-a_j)(x-b_j), 
\end{equation}
where $h(x)$ is a polynomial of the degree, $\deg h=p-q-1$, and $R^{1/2}_+(x)$ means the
value on the upper cut of
the principal sheet  of the function
$R^{1/2}(z)$ with cuts on $J$. 
The equilibrium measure
is uniquely determined by the Euler-Lagrange
conditions (see \cite{DKMVZ}): for some real constant $l$,
\begin{align}\label{qi2a}
&2\int_{\R}\log|x-s|\,d\nu_{\rm eq}(s)-V(x)=l,
\quad{\rm for}\quad x\in \cup_{j=1}^q
[a_j,b_j]\,,\\ 
&2\int_{\R}\log|x-s|\,d\nu_{\rm eq}(s)-V(x)\le l,\quad{\rm for}\quad x\in
\R\setminus\cup_{j=1}^q [a_j,b_j]\,. 
\end{align}
Equations (\ref{qi2}), (\ref{qi2a}) imply that 
\begin{equation}\label{qi3}
\om(z)=\frac{V'(z)}{2}-\frac{h(z)R^{1/2}(z)}{2},
\end{equation}
where
\begin{equation}\label{qi4}
\om(z)\equiv\int_{J} \frac {\rho(x)\,dx}{z-x}=z^{-1}+O(z^{-2}),
\quad z\to\infty\,.
\end{equation}
In addition, (\ref{qi2a}) implies that
\begin{equation}\label{qi4a}
\int_{b_j}^{a_{j+1}}\frac{h(x)R^{1/2}(x)}{2}\,dx=0\,,\quad
j=1,\ldots,q-1\,,
\end{equation}
which shows that $h(x)$ has at least one zero on each interval
$b_j<x<a_{j+1}\,;\,$ $j=1,\ldots,q-1$. From (\ref{qi3}) we obtain that
\begin{equation}\label{qi3a}
V'(z)={\rm Pol}\,\left[ h(z)R^{1/2}(z)\right],\quad \underset{z=\infty}{\rm
  Res}\, \left[h(z)R^{1/2}(z)\right]=-2\,,  
\end{equation}
and
\begin{equation}\label{qi3b}
h(z)={\rm Pol}\,\left[\frac{V'(z)}{R^{1/2}(z)}\right]\,,  
\end{equation}
where Pol$\,\left[f(z)\right]$ is the polynomial part of $f(z)$ at $z=\infty$.
The latter equation expresses $h(z)$ in terms of $V(z)$ and the end-points,
$a_1,\,b_1,\ldots,a_q,\,b_q$. The end-points can be further found from 
(\ref{qi3a}), which gives $q+1$ equation on $a_1,\ldots,b_q$, and from
(\ref{qi4a}), which gives the remaining $q-1$ equation.   

The equilibrium measure $d\nu_{\rm eq}(x)$ is called {\it regular} (otherwise
{\it singular}), 
see \cite{DKMVZ}, if 
\begin{equation}\label{qi4b}
h(x)\not=0\quad {\rm for}\quad x\in \cup_{j=1}^q [a_j,b_j]
\end{equation}
and
\begin{equation}\label{qi4c}
2\int\log|x-s|\,d\nu_{\rm eq}(s)-V(x)< l,\quad{\rm for}\quad x\in
\R\setminus\cup_{j=1}^q [a_j,b_j]\,. 
\end{equation}
The polynomial $V(x)$ is called {\it critical} if the corresponding 
equilibrium mesaure $d\nu_{\rm eq}(x)$ is singular. To study the
critical behavior in a vicinity of a critical polynomial $V(x)$, 
one embeds $V(x)$ into a parametric family $V(x;t)$, $t=(t_1,\ldots,t_r)$,
so that for some $t^c$, $V(x;t^c)=V(x)$, and 
the problem is then to evaluate the asymptotics
of eigenvalue correlation functions
as $t\to t^c$. The number of parameters $r$ depends, in general,
 on the degree of degeneracy of the equilibrium measure $d\nu_{\rm eq}(x)$.

In this paper we concern with the critical behavior
for the polynomial $V(x)$ such that the corresponding
equilibrium measure
 is supported by the segment $[-2,2]$, with a density function of the form
\begin{equation}\label{cr1}
\rho(x)=Z^{-1}(x-c)^{2m}\sqrt{4-x^2}\,,
\quad Z=\int_{\R} (x-c)^{2m}\sqrt{4-x^2}\,dx\,, 
\end{equation}
where $-2<c<2$ and $m=1,2,\ldots$. 
The choice of the support segment is obviously not important,
because by a shift and a dilation one can reduce any segment to $[-2,2]$. 
The parameter $m$ determines the degree of degeneracy 
 of the equilibrium measure at $x=c$. Observe that when $c\not=0$,
density (\ref{cr1}) is not symmetric.

Our results are summarized as follows. We are interested in two
problems:
\begin{enumerate}
\item The singularity of the infinite volume free energy at the critical
point.
\item The double scaling limit, i.e. the limit of rescaled correlation functions
as simultaneously the volume goes to infinity and the parameter $t$ goes
to $t^c$, with an appropriate relation between $t-t^c$ and the volume.
\end{enumerate} 

{\it Case $m=1$. Free energy}. We evaluate the derivatives in $T$ of the 
(infinite volume) free
energy $F(T)$, where $T>0$ is the temperature, 
and we show that $F(T)$ can be analytically continued through the
critical value $T_c$ both from below and from above of $T_c$.
In addition, we show that $F(T)$ and its first two derivatives
are continuous at $T=T_c$, while the third derivative has a jump.
This proves that at $T=T_c$ the phase transition is of the third order.
It gives an extension of the result of \cite{GW} where 
the third order phase transition was shown for the case of a
symmetric critical $V(x)$ in the circular ensemble of random matrices.

{\it Case $m=1$. Double scaling limit.}  
 The key problem here is to derive a {\it uniform}
asymptotic formula for the recurrence coefficients of the
corresponding orthogonal polynomials. The double scaling
limit descibes a transition from a fixed point
behavior of the recurrence coefficients to a quasiperiodic
behavior (cf. \cite {DKMVZ} and \cite{BDE}), and the problem
is to derive   a  uniform asymptotic formula for the recurrence
coefficients in the transition region. We show that under a proper
substitution, the recurrence coefficients are expressed, with a uniform
error term, in terms of
the Hastings-McLeod solution to the Painlev\'e II differential
equation.  In the symmetric case ($c=0$) our solution reduces
to the one obtained in \cite{DSS}, \cite{PeS}. For a rigorous
proof of the double scaling asymptotics in the symmetric case see
\cite{BI2}, \cite{BDJ}. The both latter papers are based
on the Riemann-Hilbert approach, developed in \cite{FIK},
\cite {BI1}, \cite{DKMVZ}. It is worth mentioning also earlier
physical works \cite{BKa}, \cite{GM}, \cite{DS} which concern with
the double scaling limit of the Painlev\'e I type.
       
{\it General case, $m\ge 1$. Double scaling limit.} 
We derive
a hierarchy of nonlinear ordinary differential equations which
give, under a proper substitution, the double scaling limit of the recurrence
coefficients for all $m\ge 1$. The hierarchy admits a Lax pair of linear 
differential equations and it can be constructed 
in the framework of the general theory of isomonodromic deformations
\cite{IN}. Our particular hierarchy is known as the
Painlev\'e II hierarchy \cite{Kit} and it is related to
selfsimilar solutions of the mKdV equation \cite{PeS} (see also
\cite{Moo}).

\section {Critical Behavior for a Nonsymmetric Quartic Polynomial }

\setcounter{equation}{0}

Let us consider the critical quartic polynomial $V_c(x)$
such that
\begin{equation}\label{qi5}
V'_c(x) =\frac{1}{T_c}\,( x^3 - 4 c_1 x^2 + 2 c_2 x + 8 c_1), 
 \quad T_c=1+4c_1^2\,;\qquad V_c(0)=0\,,
\end{equation}
where we denote
\begin{equation}\label{qi6}
c_k = \cos{k\pi\epsilon} \,,\quad s_k = \sin{k\pi\epsilon}.
\end{equation}
This corresponds to the critical density
\begin{equation}\label{qi7}
\rho_c(x)=\frac{1}{2\pi T_c}(x-2c_1)^2\sqrt{4-x^2}.
\end{equation}
Observe that $0<\epsilon<1$ is a parameter of the problem
 which determines the location of the critical point, 
\begin{equation}\label{qi7a}
-2<2c_1= 2\cos\pi\epsilon<2\,.
\end{equation}
Equation (\ref{qi3}) reads in this case as
\begin{equation}\label{qi8}
\om(z)=\frac{V'_c(z)}{2}-\frac{(z-2c_1)^2\sqrt{z^2-4}}{2T_c}.
\end{equation}
 
The correlations between eigenvalues in the matrix model are expressed
in 
terms of orthogonal polynomials $P_n(x)=x^n+\ldots$ on the line with
respect to the weight $e^{-NV_c(x)}$ (see e.g. \cite{Meh}, \cite{TW}).
Let
\begin{equation}\label{qi9}
\psi_n(x)=\frac{1}{\sqrt h_n}\,P_n(x)e^{-NV_c(x)/2},\qquad n=0,1,\ldots,
\end{equation}
be the corresponding psi-functions, which form an orthonormal basis in 
$L^2$. They satisfy the basic recurrence relation (see e.g. \cite{Sze}),
\begin{equation}\label{qi10}
x \psi_n(x) = \gamma_{n+1} \psi_{n+1} + \beta_n \psi_n + \gamma_n
\psi_{n-1} \virg \gamma_n=\sqrt{\frac{h_n}{h_{n-1}}} ,
\end{equation}
and the differential equation,
\begin{align}\label{qi11}
\frac{1}{N}\,\psi'_n(x) + \frac{V'_c(x)}{2} \,\psi_n(x) &= \frac{n}{
  N}\frac{1}{ \gamma_n} 
\psi_{n-1} + \frac{1}{ T_c} \gamma_n \gamma_{n-1} \left( \beta_n
+\beta_{n-1}+\beta_{n-2} - 4 c_1 \right) \psi_{n-2}+ \frac{1}{ T_c}
\gamma_n \gamma_{n-1} \gamma_{n-2} \psi_{n-3} \,.
\end{align}
The compatibility condition of equations
(\ref{qi10}), (\ref{qi11})
 leads to the string equations,
\begin{align}
T_c \frac{n}{ N} \frac{1}{\gamma^2_n} & =  \gamma_n^2 + \gamma_{n-1}^2 +
\gamma_{n+1}^2 +  \beta_n^2 + \beta_{n-1}^2 + \beta_n\beta_{n-1} -4
c_1 (\beta_n + \beta_{n-1}) + 2c_2, \label{eqmotiong}\\  
0 & =  V'_c(\beta_n) + \gamma_n^2 (2\beta_n+\beta_{n-1}) +
\gamma_{n+1}^2 (2\beta_n+\beta_{n+1}) -4c_1 (\gamma_n^2 +
\gamma_{n+1}^2)\,. \label{eqmotionb} 
\end{align}
 
To study the critical asymptotics we embed $V_c(x)$ into a parametric
family of polynomials. To that end for any $T>0$ we define the
polynomial 
\begin{equation}
V(x;T)=\frac{1}{T}\,V(x)\,,
\end{equation} 
where $V(x)$ is such that
\begin{equation}\label{qi12}
V'(x) = x^3 - 4 c_1 x^2 + 2 c_2 x + 8 c_1 \,,
\quad V(0)=0\,.
\end{equation}
Then $V'_c(x)=V'(x;T_c)$. We call $T$ {\it temperature} and $T_c$
{\it critical temperature}. Denote
$\Delta T=T-T_c$.
Let  $\rho(x;T)$ be the equilibrium density
for the polynomial $V(x;T)$. Equation (\ref{qi3}) reads in this case, 
\begin{equation}\label{qi12a}
\om(z;T)\equiv\int_{J(T)} \frac
{\rho(x;T)\,dx}{z-x}=\frac{V'(z)}{2T}
-\frac{h(z;T)R^{1/2}(z;T)}{2T},
\end{equation}
where $h(z;T)$ is a monic polynomial in $z$.

{\it Free energy near the critical point}. The 
(infinite volume) free energy is defined as 
\begin{equation}\label{fe:1}
F(T)=-T\lim_{N\to\infty} \frac{1}{N^2}\,\ln Z_N(T),\quad
Z_N(T)=\int_{{\hcal}_N} 
\exp\left(-\frac{N}{T}\,\Tr V(M)\right) dM. 
\end{equation}
We will show that
at $T=T_c$, $F(T)$ is not analytic. To evaluate the type
of nonanalyticity at $T=T_c$, consider the function
\begin{equation}\label{fe:2}
F_1(T)=T^2\frac{d}{dT}\left( \frac{F(T)}{T}\right)=
\lim_{N\to\infty} \frac{1}{Z_N(T)}\,
\int_{{\hcal}_N}\frac{1}{N} \,\Tr V(M)\,
\exp\left(-\frac{N}{T}\,\Tr V(M)\right) dM. 
\end{equation}
It can be evaluated as
\begin{align}\label{fe:3}
F_1(T)&=\int_{J(T)} V(x)\rho(x;T)dx=-\frac{1}{4\pi i T}
\oint_C V(z)h(z;T)\sqrt{R(z;T)}\,dz
\nonumber\\
&=\frac{1}{2\pi i  }
\oint_C V(z)\om(z;T)\,dz,
\end{align}
where $C$ is any contour with positive orientation around $J(T)$,
the support of equilibrium measure. Observe that the
limits in (\ref{fe:1}), (\ref{fe:2}) exist for any polynomial $V(x)$,
due to the weak convergence of eigenvalue
 correlation functions (cf. \cite{Joh}). On the contrary, 
for the second derivative of
 $F(T;N)\equiv -(T/N^2)\ln Z_N(T)$ in $T$, the convergence $F''(T;N)\to
F''(T)$ does {\it not} hold if the equilibrium
measure has two cuts or more,
because of quasiperiodic oscillations of
$F''(T;N)$ as a function of $N$ (see \cite{BDE}).
 From (\ref{fe:3}) it follows that    
since $\om(z;T)$ is continuous on $C$ in $T$ at $T=T_c$,
$F_1(T)$ is continuous as well. Therefore, $F'(T)$ is {\it continuous} at
$T=T_c$. Consider $F''(T)$.

{\it Second derivative of the free energy}.
From (\ref{fe:3}),
\begin{equation}\label{fe:4}
\frac{d[TF_1(T)]}{dT}=\frac{1}{2\pi i  }
\oint_C V(z)\frac{d}{dT}[T\om(z;T)]\,dz.
\end{equation}
For $T>T_c$, the equilibrium measure corresponding to
$V(x:T)$ is supported by one cut $[a,b]$ and the equilibrium
density is written as
\begin{equation}\label{qi14}
\rho(x;T)=\frac{1}{2\pi T}[(x-c)^2+d^2]\sqrt{(b-x)(x-a)}\,, 
\end{equation}
where $a=a(T)$, $b=b(T)$, $c=c(T)$, $d=d(T)$.
In the one-cut case we have the equation,
\begin{equation}\label{fe:5}
\frac{d}{dT}[T\om(z;T)]=\frac{1}{\sqrt{(z-a)(z-b)}}\,,
\end{equation}
see Appendix below, hence
\begin{equation}\label{fe:6}
\frac{d[TF_1(T)]}{dT}=\frac{1}{2\pi i  }
\oint_C \frac{V(z)}{\sqrt{(z-a)(z-b)}}\,dz.
\end{equation}
This implies that
\begin{equation}\label{fe:7}
\left.\frac{d[TF_1(T)]}{dT}\right|_{T=T_c^+}=
\frac{1}{2\pi i  }
\oint_C \frac{V(z)}{\sqrt{z^2-4}}\,dz.
\end{equation}
We should mention here another useful formula valid
for $T\ge T_c$ (see \cite{DGZ}, \cite{Eyn}):
\begin{equation}\label{fe:7a}
\frac{d^2[TF(T)]}{dT^2}=2\ln \frac{b-a}{4}\,.
\end{equation}

For $T<T_c$, the equilibrium measure corresponding to
$V(x)$ is supported by two cuts $[a_1,b_1]$
and $[a_2,b_2]$. The equilibrium
density is written in this case as
\begin{equation}\label{qi16}
\rho(x;T)=\frac{1}{2\pi T}(x-c)\sqrt{(b_1-x)(x-a_1)(b_2-x)(x-a_2)}\,,
\end{equation}
where where $a_1,\,b_1,\,a_2,\,b_2,\,c$ depend on $T$ and
$b_1<c<a_2$. In the two-cut case we have the equation,
\begin{equation}\label{fe:8}
\frac{d}{dT}[T\om(z;T)]
=\frac{z-x_0}{\sqrt{(z-a_1)(z-b_1)(z-a_2)(z-b_2)}}\,, 
\end{equation}
where $x_0=x_0(T)$, 
$b_1<x_0<a_2$, is determined from the condition that
\begin{equation}\label{fe:9}
\int_{b_1}^{a_2}\frac{x-x_0}
{\sqrt{(x-a_1)(x-b_1)(x-a_2)(x-b_2)}}\,dx=0\,, 
\end{equation}
see Appendix below, hence
\begin{equation}\label{fe:9a}
\frac{d[TF_1(T)]}{dT}=\frac{1}{2\pi i  }
\oint_C \frac{V(z)
(z-x_0)}{\sqrt{(z-a_1)(z-b_1)(z-a_2)(z-b_2)}}\,dz.
\end{equation}
We have that $a_2=b_1=x_0=2c_1$ at $T=T_c^-$, hence
\begin{equation}\label{fe:10}
\left.\frac{d[TF_1(T)]}{dT}\right|_{T=T_c^-}=
\frac{1}{2\pi i  }
\oint_C \frac{V(z)}{\sqrt{z^2-4}}\,dz.
\end{equation}
Thus, 
\begin{equation}\label{fe:11}
\left.\frac{d[TF_1(T)]}{dT}\right|_{T=T_c^-}=
\left.\frac{d[TF_1(T)]}{dT}\right|_{T=T_c^+}\,,
\end{equation}
so that $F''(T)$ is {\it continuous} at $T=T_c$. Consider now $F'''(T)$.

{\it Third derivative of the free energy.}
In the one-cut case we have that
\begin{equation}\label{qi14a}
\frac{d}{d T}[(x-c)^2+d^2]\sqrt{(b-x)(x-a)}
=-\frac{2}{\sqrt{(b-x)(x-a)}}
\end{equation}
and
\begin{equation}\label{qi14b}
\frac{da}{d T}
=\frac{4}{h(a)(a-b)}\,,\quad
\frac{db}{d T}
=\frac{4}{h(b)(b-a)}\,;
\qquad h(x)=(x-c)^2+d^2\,,
\end{equation}
see Appendix below.
From (\ref{qi3b}) we find that
\begin{align}\label{qi14c}
c&=2c_1-\frac{a+b}{4}\,,\qquad
d^2=\frac{5}{16}\,(a+b)^2-c_1(a+b)-\frac{1}{2}\,ab-2\,,
\end{align}
and then that
\begin{align}\label{qi14d}
\left.h(a)\right|_{a=-2,\,b=2}=4(c_1+1)^2,\qquad
\left.h(b)\right|_{a=-2,\,b=2}=4(c_1-1)^2.
\end{align}
Therefore, $a(T)$ and $b(T)$ are analytic at $T=T_c^+$, as a solution of 
system (\ref{qi14b}) with analytic coefficients.
Equation (\ref{fe:6}) implies that $F_1(T)$, 
and hence $F(T)$, are {\it analytic} at $T=T_c^+$.
From (\ref{qi14b}) and (\ref{qi14c}) we obtain  
\begin{equation}\label{qi15}
\left.\frac{da}{dT}\right|_{T=T_c^+}
=-\frac{1}{4(1+c_1)^2}\,,
\qquad \left.\frac{db}{dT}\right|_{T=T_c^+}
=\frac{1}{4(1-c_1)^2}\,.
\end{equation}

The analyticity at $T=T_c^-$ is more difficult.
In the two-cut case we have that
\begin{equation}\label{trick1}
\frac{d}{dT}(z-c) \sqrt{(z-a_1)(z-b_1)(z-a_2)(z-b_2)}=
-\frac{2(z-x_0)}{\sqrt{(z-a_1)(z-b_1)(z-a_2)(z-b_2)}}\,. 
\end{equation}
where $b_1<x_0<a_2$ solves equation (\ref{fe:9}), and
\begin{equation}\label{trick3}
\frac{da_1}{dT}=\frac{4(a_1-x_0)}{(a_1-c)
(a_1-b_1)(a_1-a_2)(a_1-b_2)}\,,\qquad
\frac{db_2}{dT}=\frac{4(b_2-x_0)}{(b_2-c)
(b_2-a_1)(b_2-b_1)(b_2-a_2)}\,,  
\end{equation}
see Appendix below. At $T=T_c^-$ this gives that
\begin{equation}\label{trick4}
\left.\frac{da_1}{dT}\right|_{T=T_c^-}
=-\frac{1}{4(1+c_1)^2}\,,\qquad
\left.\frac{db_2}{dT}\right|_{T=T_c^-}
=\frac{1}{4(1-c_1)^2}\,.
\end{equation}
Define $d$ and $\de$ such that
\begin{equation}\label{qi17a}
b_1=c-d+\de,\quad a_2=c+d+\de\,.
\end{equation}
Then $d,\,\de\to 0$ as $T\to T_c^-$ and 
from (\ref{qi4a}),
\begin{align}\label{qi17b}
0&=\int_{c-d+\de}^{c+d+\de}(x-c)\sqrt{(c+d+\de-x)(x-c+d-\de)}\,
\sqrt{(x-a_1)(b_2-x)}\,dx\nonumber\\
&=\int_{-d}^d (u+\de)\sqrt{d^2-u^2}\,\sqrt{(u+c-a_1+\de)
(b_2-c-\de-u)}\,du\nonumber\\
&=d^3\int_{-1}^1 \left(v+\frac{\de}{d}\right)
\sqrt{1-v^2}\,\sqrt{(vd+c-a_1+\de)
(b_2-c-\de-vd)}\,dv\nonumber\\
&=d^3\sqrt{(c-a_1)(b_2-c)}\,
\left[\frac{d}{2}\, \frac{\,(a_1+b_2-2c)}{(c-a_1)(b_2-c)}\,
\int_{-1}^1 v^2
\sqrt{1-v^2}\,dv+\frac{\de}{d}\,\int_{-1}^1 
\sqrt{1-v^2}\,dv\right.
\nonumber\\
&\left.+O\left(d^3+\frac{\de^2}{d}\right)\right]=d^3\sqrt{(c-a_1)(b_2-c)}\,
\left[\frac{\pi}{16}\, \frac{\,(a_1+b_2-2c)}{(c-a_1)(b_2-c)}\,d
+\frac{\pi}{2}\,\frac{\de}{d}+O\left(d^3+\frac{\de^2}{d}\right)\right]\,,
\end{align}
hence
\begin{equation}\label{qi17c}
\de=-\frac{1}{8}\, \frac{\,(a_1+b_2-2c)}{(c-a_1)(b_2-c)}\,d^2
+O(d^4)\,.
\end{equation}
By the implicit function theorem, $\de$ is an analytic function of
$d^2$ at $d=0$.
Equation (\ref{qi3b}) gives that
\begin{align}\label{trick5}
c&=2c_1-\frac{a_1+b_2}{4}-\frac{1}{2}\,\de\,,\nonumber\\
d^2&=2(1+c_1)da_1-2(1-c_1)db_2-\frac{1}{2}\,\de^2
-\frac{5}{8}\,(da_1^2+db_2^2)-\frac{1}{4}\,da_1db_2\,;
\nonumber\\
 da_1&\equiv a_1+2,\quad db_2\equiv -2+b_2\,.
\end{align}
Therefore, from (\ref{trick4}) and (\ref{qi17c}) we obtain that as $T\to
T_c^-$, 
\begin{align}\label{qi17}
d^2&=-\frac{1}{s_1^2}\DT+O(\DT^2),\qquad \de
=-\frac{c_1}{8
  s_1^4}\,\DT+O(\DT^2); 
\nonumber\\
c&=2c_1-\frac{3c_1}{16s_1^4}\,\Delta T+O(\Delta T^2)\,,
\qquad
\frac{b_1+a_2}{2}=2c_1-\frac{5c_1}{16s_1^4}\,\Delta T+O(\Delta T^2)\,.
\end{align}
Define now $\de_0$ such that
\begin{equation}\label{trick5a}
b_1=x_0-d+\de_0,\quad a_2=x_0+d+\de_0\,.
\end{equation}
Then, similar to (\ref{qi17b}), we derive  from (\ref{fe:9}) that
\begin{align}\label{trick6}
0&=\int_{x_0-d+\de_0}^{x_0+d+\de_0}
\frac{x-x_0}{\sqrt{(x_0+d+\de_0-x)(x-x_0+d-\de_0)}\,
\sqrt{(x-a_1)(b_2-x)}}\,dx\nonumber\\
&
=\frac{d}{\sqrt{(c-a_1)(b_2-c)}}\,
\left[-\frac{\pi}{2}\, \frac{\,(a_1+b_2-2c)}{(c-a_1)(b_2-c)}\,d
+\pi\,\frac{\de_0}{d}+O\left(d^3+\frac{\de_0^2}{d}\right)\right]\,,
\end{align}
hence
\begin{equation}\label{trick7}
\de_0=\frac{1}{2}\, \frac{\,(a_1+b_2-2c)}{(c-a_1)(b_2-c)}\,d^2
+O\left(d^4\right)\,,
\end{equation}
so that as $T\to T_c^-$,
\begin{equation}\label{trick8}
\de_0
=\frac{c_1}{2\,s_1^4}\,\De T+O\left(\DT^2\right)\,.
\end{equation}
By the implicit function theorem $\de_0$ is an analytic function of
$d^2$ at $d=0$. 
Using (\ref{qi17}) we obtain that
\begin{equation}\label{trick9}
x_0
=2c_1-\frac{13c_1}{16\,s_1^4}\,\De T+O\left(\DT^2\right)\,.
\end{equation}
By (\ref{qi17a}) and  (\ref{trick5a}),
\begin{equation}\label{trick10}
x_0=c+\de-\de_0,
\end{equation}
hence equations (\ref{trick3}) can be written as
\begin{align}\label{trick3a}
&\frac{da_1}{dT}=\frac{4(a_1-c-\de+\de_0)}{(a_1-c)
(d^2-(a_1-c-\de)^2)(a_1-b_2)}\,,\qquad
\frac{db_2}{dT}=\frac{4(b_2-c-\de+\de_0)}{(b_2-c)
(d^2-(b_2-c-\de)^2)(b_2-a_2)}\,.  
\end{align}
Observe that $d^2$ is an analytic function of $a_1$, $b_2$,
and $\de$, $\de_0$ are analytic functions of $d^2$. This gives
the right hand side in (\ref{trick3a}) as analytic functions of $a_1$, 
$b_2$ and hence $a_1$, $b_2$ are analytic as functions of $T$ at
$T=T_c^-$. Set 
\begin{equation}\label{m}
m=\frac{b_1+a_2}{2}\,.
\end{equation}
 Then $b_1=m-d$, $a_2=m+d$, hence,
by (\ref{fe:9a})
\begin{equation}\label{fe:9n}
\frac{d[TF_1(T)]}{dT}=\frac{1}{2\pi i  }
\oint_C \frac{V(z)
(z-x_0)}{\sqrt{(z-a_1)(z^2-2mz+m^2-d^2)(z-b_2)}}\,dz.
\end{equation}
Since $x_0$, $m$ and $d^2$ are analytic in $T$ at $T=T_c^-$, we obtain
that $F(T)$ is {\it analytic} at $T=T_c^-$.

By (\ref{fe:6}),
\begin{equation}\label{fe:12}
\left.\frac{d^2[TF_1(T)]}{dT^2}\right|_{T=T_c^+}=\frac{1}{2\pi i  }
\oint_C V(z)\,\frac{d}{dT}\left.\left(
\frac{1}{\sqrt{(z-a)(z-b)}}\right)\right|_{T=T_c^+}\,dz\,,
\end{equation}
and by (\ref{fe:9n}),
\begin{align}\label{fe:13}
\left.\frac{d^2[TF_1(T)]}{dT^2}\right|_{T=T_c^-}
&=\frac{1}{2\pi i  }
\oint_C V(z)\left[\frac{d}{dT}\left(
\frac{1}{\sqrt{(z-a_1)(z-b_2)}}\right)
\frac{(z-x_0)}{\sqrt{z^2-2mz+m^2-d^2}}\right.
\nonumber\\
&\left.\left.+\frac{1}{\sqrt{(z-a_1)(z-b_2)}}\,
\frac{d}{dT}\left(\frac{(z-x_0)}{\sqrt{z^2-2mz+m^2-d^2}}
\right)\right]\right|_{T=T_c^-}\,dz\,.
\end{align}
Observe that by (\ref{qi15}), (\ref{trick4}),
\begin{equation}\label{fe:14}
\left.\frac{da}{dT}\right|_{T=T_c^+}=
\left.\frac{da_1}{dT}\right|_{T=T_c^-}=
-\frac{1}{4(1+c_1)^2}\,;\qquad
\left.\frac{db}{dT}\right|_{T=T_c^+}=
\left.\frac{db_2}{dT}\right|_{T=T_c^-}=
\frac{1}{4(1-c_1)^2}\,,
\end{equation}
hence
\begin{align}\label{fe:15}
\left.\frac{d^2[TF_1(T)]}{dT^2}\right|_{T=T_c^+}
&-\left.\frac{d^2[TF_1(T)]}{dT^2}\right|_{T=T_c^-}
\nonumber\\
&=-\frac{1}{2\pi i  }\left.
\oint_C V(z)\frac{1}{\sqrt{z^2-4}}\,
\frac{d}{dT}\left(\frac{(z-x_0)}{\sqrt{z^2-2mz+m^2-d^2}}
\right)\right|_{T=T_c^-}\,dz\,.
\end{align}
From (\ref{qi17}) and (\ref{trick9}) we find that
\begin{equation}\label{fe:16}
\left.\frac{d}{dT}\left(\frac{(z-x_0)}{\sqrt{z^2-2mz+m^2-d^2}}
\right)\right|_{T=T_c^-}=
-\frac{1-c_1z+c_1^2}{s_1^4(z-2c_1)^2}\,,
\end{equation}
hence
\begin{align}\label{fe:15a}
\left.\frac{d^2[TF_1(T)]}{dT^2}\right|_{T=T_c^+}
-\left.\frac{d^2[TF_1(T)]}{dT^2}\right|_{T=T_c^-}
&=\frac{1}{2\pi i  }
\oint_C V(z)\frac{1}{\sqrt{z^2-4}}\,
\frac{1-c_1z+c_1^2}{s_1^4(z-2c_1)^2}\,dz
\nonumber\\
&=-\frac{3+25c_1^2+2c_1^4}{s_1^4}<0\,.
\end{align}
Thus, $F(T)$ is {\it analytic} both at $T=T_c^+$
and $T=T_c^-$, and $F'''(T)$ has a {\it jump} at $T=T_c$.
Therefore, $T=T_c$ is a critical point of the {\it third
order} phase transition.

{\it Recurrence coefficients near the critical point.}
The recurrence coefficients $\ga_n$, $\be_n$ approach 
fixed values for $T>T_c$, see e.g. \cite{DKMVZ}. Namely, 
\begin{equation}\label{qi18}
\lim_{n,N\to\infty;\;\frac{n}{N}\to \frac{T}{T_c}}\ga_n=\ga(T),\quad
\lim_{n,N\to\infty;\;\frac{n}{N}\to \frac{T}{T_c}}\be_n=\be(T),
\end{equation}
where $\ga=\ga(T),$ $\be=\be(T)$ are fixed points of (\ref{eqmotiong}),
(\ref{eqmotionb}), so that
\begin{align}
T\, \frac{1}{\gamma^2} & =  3\gamma^2
 +  3\beta^2  -8c_1 \beta + 2c_2, \label{eqgfix}\\  
0 & =  V'(\beta) + 6\gamma^2 \be-8c_1 \gamma^2 \,. \label{eqbfix} 
\end{align}
The values $\ga=\ga(T),$ $\be=\be(T)$ can be expressed in terms
of the end-points of the cut as
\begin{equation}\label{qi19}
\ga=\frac{b-a}{4}\,,\quad \be=\frac{b+a}{2}\,,
\end{equation}
see e.g. \cite{DGZ}. Therefore, by (\ref{qi15}),
\begin{equation}\label{qi20}
\ga=1+\frac{1+c_1^2}{8s_1^4}\,\DT+O(\DT^2)\,,\quad 
\be=\frac{c_1}{2s_1^4}\,\DT+O(\DT^2)\,,\quad \DT\to 0^+.
\end{equation}

For $T<T_c$, the recurrence coefficients are asymptotically
quasi-periodic, see \cite{DKMVZ}. More precisely,
\begin{equation}\label{qi21}
\lim_{n,N\to\infty;\;\frac{n}{N}\to \frac{T}{T_c}}[\ga_n-\ga(\om
\,n+\f)]=0,\quad 
\lim_{n,N\to\infty;\;\frac{n}{N}\to
  \frac{T}{T_c}}[\be_n-\be(\om\,n+\f)]=0, 
\qquad (T<T_c), 
\end{equation}
where
\begin{align}\label{qi22}
\om&=\om(T)=1-\frac{1}{K}\int_{b_2}^\infty\frac{dz}{\sqrt{R(z)}}\,,
\qquad K=\int_{b_1}^{a_2}\frac{dz}{\sqrt{R(z)}}\,;
\\ R(z)&=(z-a_1)(z-b_1)(z-a_2)(z-b_2),
\nonumber
\end{align}
$\ga(x)=\ga(x;T)$, $\be(x)=\be(x;T)$ are explicit analytic even
periodic functions of period 1 in $x$, 
and $\f$ is an explicit phase, see \cite{BDE}. The extrema of $\ga(x)$
and $\be(x)$ are expressed in terms of the end-points of the cuts,
\begin{align}\label{qi22a}
&\min_x \ga(x)=\frac{b_2-a_1-(a_2-b_1)}{4},\qquad \max_x
\ga(x)=\frac{b_2-a_1+(a_2-b_1)}{4}\,,\nonumber\\ 
&\min_x \be(x)=\frac{b_2+a_1-(a_2-b_1)}{2},\qquad \max_x
\be(x)=\frac{b_2+a_1+(a_2-b_1)}{2}\,. 
\end{align}
Using (\ref{qi17}), we obtain that as $\DT\to 0^-$,
\begin{equation}\label{qi23}
K=\frac{\pi}{2 s_1}+O(\DT),\quad
\int_{b_2}^\infty\frac{dz}{\sqrt{R(z)}}=\frac{\pi(1-\ep)}{2
  s_1}+O(\DT), 
\end{equation}
hence
\begin{equation}\label{qi24}
\om=\ep+O(\DT)\,,\quad \DT\to 0^-\,.
\end{equation}
As concerns the extrema of $\ga(x)$ 
and $\be(x)$, they behave as
\begin{align}\label{qi25}
&\min_x ,\,\max_x\ga(x)=
1\pm\frac{1}{2s_1}\left(\frac{|\DT|}{2}\right)^{1/2}
+\frac{1+c_1^2}{8s_1^4}\,\Delta
T+O(|\DT|^{3/2})\,;\\ 
&\min_x,\,\max_x \be(x)=
\pm\frac{1}{s_1}\left(\frac{|\DT|}{2}\right)^{1/2}
+\frac{c_1}{2s_1^4}\,\Delta T
+O(|\DT|^{3/2})\,,\quad\DT\to 0^-.
\end{align}

\subsection{Double Scaling Limit for Recurrence Coefficients}

We considered above the case when we took first the limit
$n,\,N\to\infty$, $\frac{n}{N}\to \frac{T}{T_c}\,,$ and then the limit
$T\to T_c$. 
Here we will consider the double scaling limit, when $n,\,N\to \infty$,
$\frac{n}{N}\to 1$, with an appropriate scaling of $n-N$. 
We start with the following ansatz, which reproduces the quasiperiodic 
behavior of the recurrence coefficients:
\begin{align}
\label{ansatz:1}
\frac{n}{ N}  =  1 & + N^{-2/3} t,\\
\gamma_n^2  =  1 & +  N^{-1/3} u(t) \cos{2n\pi \epsilon}
\nonumber\\ 
& +  N^{-2/3} \left( v_0(t) + v_1(t) \cos{2n \pi \epsilon}
+ v_2(t) \cos{4n\pi \epsilon} \right) \nonumber\\ 
 & +  N^{-1} \left( w_0(t) + w_1(t) \cos{2n \pi \epsilon} +
w_2(t) \cos{4n\pi \epsilon} + w_3(t) \cos{6n\pi \epsilon}
+w_4(t)\sin{4n\pi \epsilon} \right),
\label{ansatz:2}
\\
\beta_n = 0 & +   N^{-1/3} u(t) \cos{(2n+1)\pi \epsilon}\nonumber \\
 & +  N^{-2/3} \left( \wt{v}_0(t) + \wt{v}_1(t)
\cos{(2n+1)\pi \epsilon} + \wt{v}_2(t) \cos{(4n+2)\pi \epsilon}
\right)  \nonumber \\ 
  & +  N^{-1} \left( \wt{w}_0(t) + \wt{w}_1(t)
\cos{(2n+1)\pi \epsilon} + \wt{w}_2(t) \cos{(4n+2)\pi \epsilon} +
\wt{w}_3(t) \cos{(6n+3)\pi \epsilon}\right.\nonumber\\
&\left. +\wt{w}_4(t) \sin{(4n+2)\pi \epsilon}\right) , 
\label{ansatz:3}
\end{align}
where $u(t)$, $v_0(t),\ldots,\wt w_4(t)$ are unknown functions.
We substitute the ansatz into string equations
(\ref{eqmotiong}), (\ref{eqmotionb}) and equate terms of the same
order.

{\bf Order $N^{-1/3}$}.
Our ansatz is automatically satisfied at this order.

{\bf Order $N^{-2/3}$}. We obtain from (\ref{eqmotiong}),
(\ref{eqmotionb}) that
\begin{align}
\label{2/3:a}
\begin{pmatrix}
c_1^2+1 & - 2c_1 \\ - 2c_1 & c_1^2+1
\end{pmatrix}
\begin{pmatrix}
v_0 \\ \wt{v}_0
\end{pmatrix}
&=
\frac{u^2}{4}\begin{pmatrix}
c_1^2 \\ -c_1
\end{pmatrix}
+\frac{T_c t}{4}
\begin{pmatrix}
1 \\ 0
\end{pmatrix},\\
\label{2/3:b}
\begin{pmatrix}
1 & -1 \\ -1 & 1
\end{pmatrix}
\begin{pmatrix}
v_1 \\ \wt v_1
\end{pmatrix}
&=\frac{u'}{2}
\begin{pmatrix}
-1 \\ 1
\end{pmatrix},\\
\label{2/3:c}
\begin{pmatrix}
c_1^2+c_2^2 & - 2c_1 c_2 \\ - 2c_1 c_2 & c_1^2+c_2^2
\end{pmatrix}
\begin{pmatrix}
v_2 \\ \wt v_2
\end{pmatrix}
&=\frac{u^2}{4}
\begin{pmatrix}
1 \\ -c_3
\end{pmatrix}.
\end{align}
By solving these equations we obtain that
\begin{equation}\label{2/3:d}
v_0=-\frac{c_1^2}{ 4 s_1^2} u^2 + \frac{1+c_1^2}{ 4 s_1^4}tT_c, \qquad 
\wt{v}_0 =  -\frac{c_1 }{ 4 s_1^2} u^2 + tT_c\frac{c_1}{ 2 s_1^4}\,,
\end{equation}
and
\begin{equation}\label{2/3:e}
\wt{v}_1-v_1 = \frac{1}{ 2} u',\qquad
v_2=\frac{u^2}{ 4 s_1^2},\qquad \wt{v}_2 = \frac{c_1 u^2}{ 4 s_1^2}\,.
\end{equation}

{\bf Order $N^{-1}$}. We obtain from (\ref{eqmotiong}),
(\ref{eqmotionb}) and (\ref{2/3:d}), (\ref{2/3:e}) that 
\begin{align}
\label{1:ac}
\begin{pmatrix}
c_1^2+1 & - 2c_1 \\ - 2c_1 & c_1^2+1
\end{pmatrix}
\begin{pmatrix}
w_0 \\ \wt{w}_0
\end{pmatrix}
&=\frac{uu'}{4}
\begin{pmatrix}
0 \\ -c_1
\end{pmatrix}
+\frac{uv_1}{2}
\begin{pmatrix}
c_1^2 \\ -c_1
\end{pmatrix}
+
c_1\begin{pmatrix}
-\wt v'_0 \\ v'_0 
\end{pmatrix},\\
\label{1:bc}
8c_1^2\begin{pmatrix}
1 & -1 \\ -1 & 1
\end{pmatrix}
\begin{pmatrix}
w_1 \\ \wt w_1
\end{pmatrix}
&=\frac{u^3}{2s_1^2}
\begin{pmatrix}
-1 \\ c_2
\end{pmatrix}
+\frac{T_ctu}{2s_1^4}
\begin{pmatrix}
2c_1^4+3c_1^2-1 \\ -2c_1^4-c_1^2-1
\end{pmatrix}
+u''
\begin{pmatrix}
-c_2 \\ 1
\end{pmatrix}
+v_1'\begin{pmatrix}
-4c_1^2 \\ 4c_1^2
\end{pmatrix},\\
\label{1:dc}
\begin{pmatrix}
c_1^2+c_2^2 & - 2c_1 c_2 \\ - 2c_1 c_2 & c_1^2+c_2^2
\end{pmatrix}
\begin{pmatrix}
w_2 \\ \wt w_2
\end{pmatrix}
&=\frac{uv_1}{2}
\begin{pmatrix}
1 \\ -c_3
\end{pmatrix}
+\frac{uu'}{4s_1^2}
\begin{pmatrix}
-2c_1^2c_2 \\ 4c_1^5-3c_1^3+c_1
\end{pmatrix},\\
\label{1:ec}
\begin{pmatrix}
c_1^2+c_3^2 & - 2c_1 c_3 \\ - 2c_1 c_3 & c_1^2+c_3^2
\end{pmatrix}
\begin{pmatrix}
w_3 \\ \wt w_3
\end{pmatrix}
&=\frac{u^3c_1^2}{8s_1^2}
\begin{pmatrix}
-c_2+2 \\ -2c_1c_3+1
\end{pmatrix},\\
\label{1:fc}
\begin{pmatrix}
c_1^2+c_2^2 & - 2c_1 c_2 \\ - 2c_1 c_2 & c_1^2+c_2^2
\end{pmatrix}
\begin{pmatrix}
w_4 \\ \wt w_4
\end{pmatrix}
&=\frac{uu'}{8}
\begin{pmatrix}
-2s_2 \\ (4c_1^4-3c_1^2-1)/s_1
\end{pmatrix}
\end{align}
(we did symbolic calculations with MAPLE). 
Consider equation (\ref{1:bc}). The matrix on the left in this
equation is degenerate, 
hence we have the compatibility condition,
\begin{equation}\label{compat:1}
\boxed{2s_1^2u''=u^3+\frac{T_c}{s_1^2}\,tu,}
\end{equation}
which is the Painlev\'e II equation. When $\eps=1/2$ it reduces to 
$2u''=u^3+tu$. The function $u(t)$ behaves as
\begin{equation}\label{compat:5}
u(t) \mathop{\sim}_{t\to -\infty} \frac{1}{s_1} \,\sqrt{-T_c\, t\,}\,,
\qquad
u(t) \mathop{\sim}_{t\to +\infty} {\rm Ai}\,(\kappa\,t),\qquad
\kappa=\left(\frac{T_c}{2s_1^4}\right)^{1/3}. 
\end{equation}
Here and in what follows we use the following notations:
$f(x)\sim g(x)$ as $x\to a$ means that $\lim_{x\to
  a}\frac{f(x)}{g(x)}=1$, and $f(x)\approx g(x)$ as $x\to a$ means
that $\lim_{x\to 
  a}[f(x)-g(x)]=0.$

\subsection {Scaled Differential Equations at the Critical Point}

Equations  (\ref{qi10}), (\ref{qi11}) can be used to derive a closed
system of differential equations,
\begin{equation}\label{diff:1}
\frac{T_c}{N}\,\frac{d}{dx}
\begin{pmatrix}
\psi_n \\ \psi_{n-1}
\end{pmatrix}
=
\begin{pmatrix}
-\frac{TV'(x)}{2}-\ga_n^2A_n(x) & \ga_nB_n(x) \\
-\ga_n B_{n-1}(x) & \frac{TV'(x)}{2}+\ga_n^2A_n(x)
\end{pmatrix} 
\begin{pmatrix}
\psi_n \\ \psi_{n-1}
\end{pmatrix},
\end{equation}
where
\begin{align}
\label{diff:2}
A_n(x)&=x-4c_1+\be_n+\be_{n-1},\\
\label{diff:3}
B_n(x)&=x^2 + x (\beta_n-4c_1) + \beta_n^2 - 4c_1 \beta_n +2c_2
+\gamma_n^2+\gamma_{n+1}^2. 
\end{align}
To derive a scaled system at the critical point $x=2c_1$ we set
\begin{equation}\label{diff:4}
x=2c_1+yN^{-1/3}.
\end{equation}
Then
\begin{align}
\label{diff:5}
\frac{T_cV'(x)}{2}+\ga_n^2A_n(x)
&=
2c_1(1-\ga_n^2)+\ga_n^2(\be_n+\be_{n-1})+(\ga_n^2-1)yN^{-1/3}
+c_1y^2N^{-2/3}+\frac{y^3N^{-1}}{2}\,,\\
B_n(x)&=y^2 + \beta_n y + \gamma_n^2 +\gamma_{n+1}^2 -2 -2c_1 \beta_n + \beta_n^2.
\end{align}
Substituting ansatz (\ref{ansatz:1})-(\ref{ansatz:3}) we obtain that
\begin{align}
\label{diff:6}
\frac{T_cV'(x)}{2}+\ga_n^2A_n(x)
&=
N^{-2/3} 
\left( c_1 y^2 +\frac{c_1 u^2}{2} +\frac{c_1 T_ct}{2 s_1^2}  
+yu\cos 2n\pi\epsilon - s_1 u' \sin 2n\pi\epsilon \right)
+O(N^{-1})\,,\\
\ga_n B_n(x)
&=
N^{-2/3}\left( y^2 + \frac{u^2}{2} + \frac{T_ct}{2s_1^2}  
+ yu\cos{(2n+1)\pi\epsilon} + s_1 u' \sin{(2n+1)\pi\epsilon}\right)
+O(N^{-1})\,,\nonumber\\
\gamma_n B_{n-1} &=
 N^{-2/3}\left( y^2 + \frac{u^2}{2} + \frac{T_ct}{2 s_1^2}
+ yu\cos{(2n-1)\pi\epsilon} + s_1 u' \sin{(2n-1)\pi\epsilon}\right)
+O(N^{-1})\,.\nonumber
\end{align}
Thus, system (\ref{diff:1}) reduces to 
\begin{equation}\label{diff:7}
T_c\,\frac{d}{dy}
\begin{pmatrix}
\psi_n \\ \psi_{n-1}
\end{pmatrix}
=
\begin{pmatrix}
a_{11}(y) & a_{12}(y) \\
a_{21}(y) & a_{22}(y)
\end{pmatrix} 
\begin{pmatrix}
\psi_n \\ \psi_{n-1}
\end{pmatrix},
\end{equation}
where up to $O(N^{-1/3})$,
\begin{align}
a_{11}(y)&=-c_1\left(y^2+\frac{u^2}{2}+\frac{T_ct}{2s_1^2}\right)
-yu\cos 2n\pi\epsilon+s_1u'\sin 2n\pi\epsilon,\\
a_{12}(y)&=y^2+\frac{u^2}{2}+\frac{T_ct}{2s_1^2}
+yu\cos (2n+1)\pi\epsilon+s_1u'\sin (2n+1)\pi\epsilon,\\
a_{21}(y)&=-\left(y^2+\frac{u^2}{2}+\frac{T_ct}{2s_1^2}\right)
-yu\cos (2n-1)\pi\epsilon-s_1u'\sin (2n-1)\pi\epsilon,\\
a_{22}(y)&=c_1\left(y^2+\frac{u^2}{2}+\frac{T_ct}{2s_1^2}\right)
+yu\cos 2n\pi\epsilon-s_1u'\sin 2n\pi\epsilon.
\end{align}
When $\epsilon=1/2$, this simplifies to
\begin{align}
a_{11}(y)&=-a_{22}(y)=-(-1)^nyu,\\
a_{12}(y)&=y^2+\frac{u^2+t}{2}+(-1)^nu',\\
a_{21}(y)&=-\left(y^2+\frac{u^2+t}{2}\right)+(-1)^nu'.
\end{align}
Under the substitution
\begin{align}\label{rec4:1}
\psi_n(y)&=\cos\left(n+\frac{1}{2}\right)\pi\epsilon\,f(y)
-\sin\left(n+\frac{1}{2}\right)\pi\epsilon\,g(y),\\
\psi_{n-1}(y)&=\cos\left(n-\frac{1}{2}\right)\pi\epsilon\,f(y)
-\sin\left(n-\frac{1}{2}\right)\pi\epsilon\,g(y),
\end{align}
system (\ref{diff:7}) reduces, up to $O(N^{-1/3})$, to
\begin{equation}\label{diff:8}
\boxed{\frac{T_c}{s_1} 
\,\frac{d}{dy}\begin{pmatrix}
f \\ g
\end{pmatrix}
 = \begin{pmatrix}
 s_1 u'   &  \left(y^2 + \frac{u^2}{2}
 + \frac{Tt}{2s_1^2}\right)  + yu   \\
 -\left(y^2 + \frac{u^2}{2}
 + \frac{Tt}{2s_1^2}\right)  + yu     &  -s_1 u' 
\end{pmatrix}
\begin{pmatrix}
f \\ g
\end{pmatrix},}
\end{equation}
the differential $\psi$-equation for Painlev\'e II equation (\ref{compat:1}).

\subsection{Universal Kernel}

To eliminate dependence on $\epsilon$ consider new variables $\tilde t$, $\tilde u$ and $\tilde y$
such that 
\begin{equation}\label{uk:1}
t=\left(\frac{2s_1^4}{T_c}\right)^{1/3}\tilde t ,\quad u=\left(\frac{4T_c}{s_1}\right)^{1/3}\tilde u,
\quad y=\left(\frac{4T_c}{s_1}\right)^{1/3}\tilde y.
\end{equation}
Then equations (\ref{compat:1}) and (\ref{diff:8}) reduce to
\begin{equation}\label{uk:2}
\tilde u''=\tilde t\, \tilde u+2\tilde u^3,\quad (\,{}')=\frac{d}{d\tilde t}\,,
\end{equation}
and 
\begin{equation}\label{uk:3} 
\frac{d}{d\tilde y}\begin{pmatrix}
f \\ g
\end{pmatrix}
 = \begin{pmatrix}
 2 \tilde u'   &  \left(4\tilde y^2 + 2\tilde u^2
 + \tilde t\right)  + 4\tilde y\tilde u   \\
 -\left(4\tilde y^2 + 2\tilde u^2
 + \tilde t\right)  + 4\tilde y\tilde u     &  -2 \tilde u' 
\end{pmatrix}
\begin{pmatrix}
f \\ g
\end{pmatrix}.
\end{equation}
Equations (\ref{uk:1}) give the scaling as
\begin{align}\label{uk:4}
\frac{n}{N}&=1+N^{-2/3}\left(\frac{2s_1^4}{T_c}\right)^{1/3}\tilde t ,\\
\ga_n^2&=1+N^{-1/3}\left(\frac{4T_c}{s_1}\right)^{1/3}\tilde u\,
\cos 2n\pi\epsilon+O(N^{-2/3}),\\
\be_n&=0+N^{-1/3}\left(\frac{4T_c}{s_1}\right)^{1/3}\tilde u\,
\cos (2n+1)\pi\epsilon+O(N^{-2/3}),\\
x&=2c_1+N^{-1/3}\left(\frac{4T_c}{s_1}\right)^{1/3}\tilde y\,.
\end{align}
The Dyson integral kernel for the double scaling limit correlation
functions is then: 
\begin{equation}\label{uk:8}
\boxed{K(\tilde y_1,\tilde y_2) =  \frac{ f(\tilde y_1)g(\tilde y_2) 
- g(\tilde y_1)f(\tilde y_2)}{\tilde y_1-\tilde y_2}\,.}
\end{equation}

\section{Nonlinear Hierarchy}

\setcounter{equation}{0}

\subsection {Basic Ansatz}

For $m=1,2,\ldots$, we consider the model critical density
\begin{equation}\label{ba:1}
\rho(x)=\frac{1}{2\pi T_c}(x-2c_1)^{2m}\sqrt{4-x^2},
\end{equation}
where 
\begin{equation}\label{ba:2}
T_c=\frac{1}{2\pi }\int_{-2}^2(x-2c_1)^{2m}\sqrt{4-x^2}\,dx.
\end{equation}
The corresponding polynomial $V(x)$ is such that
\begin{equation}\label{ba:3}
V'(x)=\frac{1} {T_c}\,{\rm Pol}\,\left[(x-2c_1)^{2m}\sqrt{4-x^2}\right],
\end{equation}
where Pol$[f(x)]$ means a polynomial part of a function $f(x)$ at infinity.
In particular,
\begin{align}\label{ba:4}
m=1:\quad V'(x)&=\frac{1}{T_c}\left[x^3-4c_1x^2+2c_2x+8c_1\right],\quad
T_c=1+4c_1^2\,;\\ 
m=2:\quad V'(x)&=\frac{1}{T_c}\left[x^5-8c_1x^4+(-2+24c_1^2)x^3-16c_1c_2x^2
+(-2+16c_1^4-48c_1^2)x\right.\nonumber\\
&\left.+16(c_1+4c_1^3)\right],
\quad T_c=2+24c_1^2+16c_1^4\,,
\end{align}
and so on. In fact, our considerations will be very general and (\ref{ba:1})
is only an example. They hold for any density (\ref{qi2}) 
which satisfies regularity conditions (\ref{qi4b}), (\ref{qi4c})
everywhere except one point $c$ lying strictly inside one of the cuts,
and such that as $z\to c$, $h(z)\sim C(x-c)^{2m}$, $C\not=0$.

In the double scaling limit we define variables $K$, $t$ and $y$ as
\begin{equation}
K=N^{-1/(2m+1)}\,,\qquad \frac{n}{N} = 1 + K^{2m} s_1 t \,,\qquad x =
2 c_1 + 2 K y\,. 
\end{equation}
Our ansatz for the orthogonal polynomials is the following:
\begin{align}\label{ba:5}
\psi(n,x) &= \cos{(n+1/2)\pi\eps}\, f(t,y) - \sin{(n+1/2)\pi\eps}\,
g(t,y)\nonumber\\
&+K\left[\cos{(n+1/2)\pi\eps}\, f_1(t,y) - \sin{(n+1/2)\pi\eps}\,
g_1(t,y)\right.\nonumber\\
&+\left.
\cos{3(n+1/2)\pi\eps}\,\wt f(t,y) - \sin{3(n+1/2)\pi\eps}\,
\wt g(t,y)\right] +O(K^2), 
\end{align} 
[cf. (\ref{rec4:1})], and for the recurrence coefficients,
\begin{equation}\label{ba:6}
\gamma_n = 1 + K u(t)\,\cos{2n\pi\eps}+O(K^2) \,,\qquad \beta_n = 2 K
u(t)\, 
\cos{(2n+1)\pi\eps}+O(K^2)\, 
\end{equation}
[cf. (\ref{ansatz:2}), (\ref{ansatz:3})]. See also 
the work \cite{PeS} where an 
intimately related ansatz for the recurrence coefficients was
suggested in the case of
 a {\it symmetric} potential $V(x)$ in the circular ensemble.
We substitute ansatz (\ref{ba:5}), 
(\ref{ba:6}) into the 3-terms recursion relation,
\begin{equation}\label{ba:7}
x \psi(n,x) = \gamma_{n+1} \psi(n+1,x) + \beta_n \psi(n,x) + \gamma_n
\psi(n-1,x) 
\end{equation}
and in the first order in $K$ we obtain two systems of equations,
\begin{equation}\label{ba:8}
\partial_t
\begin{pmatrix}
f(t,y) \\
g(t,y)
\end{pmatrix}
=L\begin{pmatrix}
f(t,y) \\
g(t,y)
\end{pmatrix},\qquad
L=\begin{pmatrix}
0 & y +u(t) \\
-y+u(t) & 0
\end{pmatrix}
\end{equation}
(at frequency 1) and
\begin{equation}\label{ba:9}
\begin{pmatrix}
\wt f(t,y) \\
\wt g(t,y)
\end{pmatrix}
=\frac{c_1u(t)}{4s_1^2}\begin{pmatrix}
f(t,y) \\
g(t,y)
\end{pmatrix}
\end{equation}
(at frequency 3).

\subsection {Differential System}

We would like to derive a differential equation in $y$,
\begin{equation}\label{ds:1}
\partial_y
\begin{pmatrix}
f(t,y) \\
g(t,y)
\end{pmatrix}
=D(t,y)\begin{pmatrix}
f(t,y) \\
g(t,y)
\end{pmatrix}.
\end{equation}
We are looking for $D(t,y)$ in the form
\begin{equation}\label{ds:2}
D(t,y)=
\begin{pmatrix}
-A(t,y) & yB(t,y)+C(t,y) \\
yB(t,y)-C(t,y) & A(t,y)
\end{pmatrix}
\end{equation}
[cf. (\ref{diff:8})], where $A$, $B$ and $C$ are even
polynomials in $y$ of the following degrees:
\begin{equation}\label{ds:3}
\deg A=2m-2,\quad\deg B=2m-2,\quad \deg C=2m.
\end{equation}
We will assume that $C$ is a monic polynomial, so that
$C=y^{2m}+\ldots$. The general case can be reduced to this one
by the change of variables, $t=\kappa \tilde t$, 
$y=\frac{\tilde y}{\kappa}\,$, $u(t)=\frac{\tilde u(\tilde t)}{\kappa}\,$,
which preserves the structure of the operator $L$ in (\ref{ba:8}).  
The consistency condition of equations (\ref{ba:8}) and (\ref{ds:1}),
\begin{equation}\label{ds:4}
[D,L]=\partial_y L-\partial_t D=
\begin{pmatrix}
0 & 1 \\
-1 & 0
\end{pmatrix}-\partial_t D,
\end{equation}
implies that
\begin{equation}\label{ds:5}
\partial_tB=2A\,,\quad
\partial_t C=1+2uA\,,\quad
\partial_t A=-2y^2B+2uC\,.
\end{equation}

$\bullet$ Example: $m=1$. According to (\ref{ds:3}),
$A=a(t)$, $B=b(t)$, $C=y^2+c(t)$.
From the last equation in (\ref{ds:5}) we obtain that $b=u$ and then
that
\begin{equation}\label{ds:6}
a=\frac{u'}{2},\quad b=u,\quad c=t+\frac{u^2}{2}+t_0,
\end{equation} 
where $t_0$ is a free constant, and
\begin{equation}\label{ds:7}
\frac{u''}{2}=u^3+2(t+t_0)u,
\end{equation} 
the Painlev\'e II equation. By changing $t+t_0$ to $t$ we can reduce
it to $t_0=0$. 

We would like to construct solutions to 
(\ref{ds:5}) for $m>1$. To that end, define recursively
functions $A_m(t,y)$, $B_m(t,y)$, $C_m(t,y)$ by the equations
\begin{align}\label{ds:8a}
C_{m+1}&=y^2C_m+f_m(u)\,,\\
\label{ds:8b}
B_{m+1}&=y^2B_m+R_m(u)\,,\\
\label{ds:8c}
A_{m+1}&=y^2A_m+\frac{1}{2}\,\partial_t R_m(u)\,,
\end{align} 
where $R_m(u)$, $f_m(u)$ solve the recursive equations
\begin{align}\label{ds:9a}
R_{m+1}(u)&=uf_m(u)-\frac{1}{4}\,\partial_{tt}R_m(u),\\
\label{ds:9b}
\partial_t f_m(u)&=u\partial_t R_m(u),\qquad f_m(0)=0,
\end{align} 
with the initial data
\begin{equation}\label{ds:10}
A_0=B_0=0,\quad C_0=1,\quad R_0(u)=u,\quad f_0(u)=\frac{u^2}{2}\,.
\end{equation} 
We solve recursively (\ref{ds:9a})-(\ref{ds:10}) as
\begin{align}\label{ds:11}
R_1(u)&=\frac{1}{2}u^3-\frac{1}{4}\,u'',\qquad f_1(u)=\frac{3}{8}u^4
-\frac{1}{4}\,uu''+\frac{1}{8}{u'}^2\,;\\
R_2(u)&=\frac{3}{8}\,u^5-\frac{5}{8}u^2u''-\frac{5}{8}\,u{u'}^2
+\frac{1}{16}\,u^{(4)},\\
f_2(u)&=\frac{5}{16}\,u^6-\frac{5}{8}\,u^3u''-\frac{5}{16}\,u^2{u'}^2+
\frac{1}{16}uu^{(4)}-\frac{1}{16}u'u'''+\frac{1}{32}{u''}^2\,;\\
R_3(u)&=\frac{5}{16}\,u^7-\frac{35}{32}u^4u''-\frac{35}{16}\,u^3{u'}^2
+\frac{7}{32}\,u^2u^{(4)}+\frac{7}{8}\,uu'u'''+\frac{21}{32}u{u''}^2
+\frac{35}{32}\,{u'}^2u''-\frac{1}{64}u^{(6)} , 
\end{align} 
and so on,
\begin{equation}\label{ds:12}
R_m(u)=\frac{(2m)!}{2^{2m}(m!)^2}\,u^{2m+1}+\ldots+\frac{(-1)^m}{2^{2m}}u^{(2m)}.
\end{equation} 
In addition,
\begin{align}\label{ds:12a}
A_1&=\frac{1}{2}\,u',\quad B_1=u,\quad C_1=y^2+\frac{1}{2}\,u^2\,;\\
A_2&=\frac{1}{2}\,u'y^2+\frac{3}{4}\,u^2u'-\frac{1}{8}\,u''',\quad
B_2=uy^2+\frac{1}{2}\,u^3-\frac{1}{4}\,u'',\\
C_2&=y^4+\frac{1}{2}\,u^2y^2+\frac{3}{8}\,u^4-\frac{1}{4}\,uu''+
\frac{1}{8}\,(u')^2\,;
\end{align} 
and so on.
It is easy to check that the functions $A_m(t,y)$,
$B_m(t,y)$, $C_m(t,y)$  defined by (\ref{ds:8a})-(\ref{ds:10})
solve the equations
\begin{equation}\label{ds:13}
\partial_tB_m=2A_m,\quad \partial_t C_m=2uA_m,\quad \partial_t A_m=
-2y^2B_m+2uC_m-2R_m(u).
\end{equation}
Indeed, by (\ref{ds:10}) it holds for $m=0$. Assume that it holds for
some $m$. Then by (\ref{ds:8a})-(\ref{ds:9b}) and (\ref{ds:13}),
\begin{align}\label{ds:13a}
\partial_t C_{m+1}&=y^2\partial_t C_m+\partial_t
f_m(u)=2y^2uA_m+u\,\partial_{t} R_m(u)=2uA_{m+1}\,,\\ 
\partial_t B_{m+1}&=y^2\partial_t B_m+\partial_t
R_m(u)=2y^2A_m+\partial_t R_m(u)=2A_{m+1}\,,\\ 
\partial_t A_{m+1}&=y^2\partial_t A_m+\frac{1}{2}\,\partial_{tt} R_m(u)
=y^2\left[-2y^2B_m+2uC_m-2R_m(u)\right]+\frac{1}{2}\,\partial_{tt}
R_m(u)\nonumber\\ 
&=-2y^2\left[y^2B_m+R_m(u)\right]+2u\left[y^2C_m+f_m(u)\right]
-2\left(uf_m(u)-\frac{1}{4}\,\partial_{tt}R_m(u)\right)
\nonumber\\
&=-2y^2B_{m+1}+2uC_{m+1}-2R_{m+1}(u)\,,
\end{align}
which proves (\ref{ds:13}) for $m+1$ and hence for all $m=0,1,2,\ldots$. 
Comparing (\ref{ds:5}) with (\ref{ds:13}) we obtain that 
\begin{equation}\label{ds:13b}
A=A_m,\quad B=B_m,\quad C=t+C_m
\end{equation}
solve equation (\ref{ds:5}), provided $u(t)$ is a solution of the
equation 
\begin{equation}\label{ds:14}
R_m(u)+tu=0\,.
\end{equation}
The sequence of equations (\ref{ds:14}) for $m=1,2,\ldots$
forms a hierarchy of ordinary differential equations
which is known as the Painlev\'e II hierarchy \cite{Kit} 
(see also \cite{Moo}, \cite{PeS}).  
We can formulate now the following result.

\begin{theo} 
Define
\begin{equation}\label{ds:15}
D_m(t,y)=
\begin{pmatrix}
-A_m(t,y) & yB_m(t,y)+C_m(t,y) \\
yB_m(t,y)-C(t,y) & +A_m(t,y)
\end{pmatrix}
\end{equation}
Then if $u(t)$ is a solution of equation (\ref{ds:14}),
then the matrix
\begin{equation}\label{ds:16}
D(t,y)=
\begin{pmatrix}
0 & t \\
-t & 0
\end{pmatrix}+D_m(t,y)
\end{equation} 
is a solution to (\ref{ds:4}). More generally,
if $t_1,\ldots,t_m$ are arbitrary constants and $u(t)$ is
a solution of the equation
\begin{equation}\label{ds:17}
\sum_{k=1}^m t_k R_k(u)+tu=0,
\end{equation}  
then the matrix
\begin{equation}\label{ds:18}
D(t,y)=
\begin{pmatrix}
0 & t \\
-t & 0
\end{pmatrix}+\sum_{k=1}^m t_kD_k(t,y)
\end{equation} 
is a solution to (\ref{ds:4}). 
\end{theo}

\begin{rem} It can be shown that (\ref{ds:18}) is a {\it general}
solution to equation (\ref{ds:4}). 
\end{rem}

The meaning of the constants $t_1,\dots,t_m$ in (\ref{ds:18}) is the
following. Observe that
the differential equation in $y$, (\ref{ds:1}) describes the double scaling
limit for a critical polynomial of degeneracy $2m$. In this case
the space of transversal fluctuations to the  manifold of
 critical polynomials has dimension $m$. The variables $t_1,\dots,t_m$
serve as coordinates in the space of transversal fluctuations,
and (\ref{ds:18}) gives the matrix describing the double scaling limit
of the recurrence coefficients
in the direction $\tau=(t_1,\ldots,t_m)$.  

\section{Conclusion}

In this paper we considered critical polynomials which violate
the regularity conditions at exactly one point, inside the
support of the equilibrium measure. It is characterized by the
degree $2m$ of degeneracy of the equilibrium density at the critical point.
Our main results are the following:
\begin{itemize}
\item When $m=1$, the infinite volume free energy exhibits the phase
transition of the third order. This extends the result of \cite{GW}
to nonsymmetric critical polynomials.
\item When $m=1$, the double scaling limit of the recurrence coefficients
is described, under a proper substitution, by the Hastings-McLeod
solution to the Painlev\'e II differential equation. Before this result
was known only for symmetric critical polynomials \cite{DSS}, \cite{PeS}
(for rigorous results see \cite{BI2}, \cite{BDJ}).
\item For $m>1$, we derive a hierarchy of ordinary differential equations
describing the double scaling limit of the recurrence coefficients.
\end{itemize}

\section{Appendix. One Useful Identity }

\setcounter{equation}{0}

Let 
$V(z;T)=\frac{V(z)}{T}\,,$
where $T>0$ is the temperature, and
\begin{align}\label{qi1a}
\mu_N(dM;T)&=Z_N(T)^{-1}\exp\left(-\frac{N}{T}\,\Tr V(M)\right) dM,
\qquad
Z_N(T)=\int_{{\hcal}_N} 
\exp\left(-\frac{N}{T}\,\Tr V(M)\right) dM\,.  
\end{align} 
Then 
$\rho(x)$ and $\om(z)$ depend on
$T$. The following identity is useful in many questions.

{\bf Proposition.} 
{\it Assume that the number of cuts does not change in a neighborhood of a
given $T>0$. Then
\begin{equation}\label{qi4y}
\frac{d}{dT}[T\om(z)]=\frac{\prod_{j=1}^{q-1}(z-x_j)}{R^{1/2}(z)}\,,
\end{equation}
where the numbers $b_j<x_j<a_{j+1}$, $j=1,\dots,q-1,$
solve the equations,
\begin{equation}\label{qi4z}
\int_{b_k}^{a_{k+1}}\frac{\prod_{j=1}^{q-1}(x-x_j)}
{R^{1/2}(x)}\,dx=0,\quad   k=1,\dots,q-1\,.
\end{equation}
The neighborhood can be one-sided, then the derivative in $T$ is also
one-sided.}

{\it Proof.} Equation (\ref{qi12a}) gives that
\begin{equation}\label{app:1}
T\om(z)=\frac{V'(z)}{2}-\frac{h(z)R^{1/2}(z)}{2}\,.
\end{equation} 
Since $V(z)$ does not depend on $T$,
\begin{equation}\label{app:2}
\frac{d}{dT}[T\om(z)]=-\frac{d}{dT}\,\frac{h(z)R^{1/2}(z)}{2}\,.
\end{equation}
The function on the right can be written as
\begin{equation}\label{app:3}
-\frac{d}{dT}\,\frac{Th(z)R^{1/2}(z)}{2}
=\frac{P(z)}{R^{1/2}(z)}\,,
\end{equation}
where $P(z)$ is a polynomial with real coefficients. Since
\begin{equation}\label{app:4}
\frac{d}{dT}[T\om(z)]=\frac{1}{z}+O(z^{-2})\,,
\end{equation}
we obtain that
\begin{equation}\label{app:5}
\frac{P(z)}{R^{1/2}(z)}
=\frac{1}{z}+O(z^{-2})\,,
\end{equation}
which shows that $P(z)=z^{q-1}+\dots$. By (\ref{qi4a}),
\begin{equation}\label{app:6}
\int_{b_j}^{a_{j+1}}\frac{h(x)R^{1/2}(x)}{2}\,dx=0\,,\quad
j=1,\ldots,q-1\,.
\end{equation}
By differentiting with respect to $T$ we obtain that
\begin{equation}\label{app:7}
\int_{b_j}^{a_{j+1}}\frac{P(x)}{R^{1/2}(x)}\,dx=0\,,\quad
j=1,\ldots,q-1\,.
\end{equation}
This is possible only if $P(x)$ has a zero in each interval
$[b_j,a_{j+1}]$. Thus, (\ref{qi4z}) is proven.

As a corollary, from (\ref{app:2}) we get that
\begin{equation}\label{app:8}
\frac{d}{dT}\,\left[h(z)R^{1/2}(z)\right]
=-\frac{2\prod_{j=1}^{q-1}(z-x_j)}{R^{1/2}(z)}\,.
\end{equation}
This implies that
\begin{equation}\label{app:9}
\frac{d}{dT}\,\ln\left[h(z)R^{1/2}(z)\right]
=-\frac{2\prod_{j=1}^{q-1}(z-x_j)}{h(z)R(z)}\,\,.
\end{equation}
Comparing the residue of the both sides at $z=a_k,b_k$ we obtain that
\begin{align}\label{app:10}
&\frac{da_k}{dT}
=\frac{4\prod_{j=1}^{q-1}(a_k-x_j)}{h(a_k)(a_k-b_k)\prod_{j\,:\,
j\not= k} [(a_k-a_j)(a_k-b_j)]}\,,\nonumber\\
&\frac{db_k}{dT}
=\frac{4\prod_{j=1}^{q-1}(a_k-x_j)}{h(b_k)(b_k-a_k)\prod_{j\,:\,
j\not= k} [(b_k-a_j)(b_k-b_j)]}\,,\quad 1\le k\le q.
\end{align}

\end{document}